# Landau level spectroscopy of valence bands in HgTe quantum wells: Effects of symmetry lowering

L. S. Bovkun,[1,2,*] A. V. Ikonnikov,[1,3] V. Ya. Aleshkin,[1,4] K. E. Spirin,[1] V. I. Gavrilenko,[1,4] N. N. Mikhailov,[5] S. A. Dvoretskii,[5] F. Teppe,[6] B. A. Piot,[2] M. Potemski,[2] M. Orlita[2,7,†]

[1]*Institute for Physics of Microstructures RAS, 603950 Nizhny Novgorod, Russia*
[2]*Laboratoire National des Champs Magnétiques Intenses, LNCMI-CNRS-UGA-UPS-INSA-EMFL, 38042 Grenoble, France*
[3]*Faculty of Physics, M.V. Lomonosov Moscow State University, 119991 Moscow, Russia*
[4]*Lobachevsky State University of Nizhny Novgorod, 603950 Nizhny Novgorod, Russia*
[5]*Institute of Semiconductor Physics, Siberian Branch RAS, 630090 Novosibirsk, Russia*
[6]*Laboratoire Charles Coulomb (L2C), UMR CNRS 5221, GIS-TERALAB, Université Montpellier II, F-34095 Montpellier, France*
[7]*Institute of Physics, Charles University, 12116 Prague, Czech Republic*

Landau level spectroscopy has been employed to probe the electronic structure of the valence band in a series of p-type HgTe/HgCdTe quantum wells with both normal and inverted ordering of bands. We find that the standard axial-symmetric 4-band Kane model, which is nowadays widely applied in physics of HgTe-based topological materials, does not fully account for the complex magneto-optical response observed in our experiments — notably, for the unexpected avoided crossings of excitations and for the appearance of transitions that are electric-dipole forbidden within this model. Nevertheless, reasonable agreement with experiments is achieved when the standard model is expanded to include effects of bulk and interface inversion asymmetries. These remove the axial symmetry, and among other, profoundly modify the shape of valence bands.

DOI:

## I. INTRODUCTION

In recent years, there has been a considerable interest in HgTe/CdHgTe quantum wells (QWs) with the narrow gap or even gapless band structures. Most notably, QWs with the inverted band structures (HgTe QWs of widths larger than $d_c \sim 6.3$ nm) were identified as the very first topological insulators, thus opening a completely new field for current condensed matter physics [1,2]. Among other recent achievements on HgTe/CdHgTe heterostructures, one may mention the realization of stimulated emission (due to interband recombination of electrons and holes) demonstrated at wavelengths up to 20 μm [3,4]. This is possible thanks to a fairly low electron-hole asymmetry of the band structure in narrow gap HgTe/CdHgTe QWs, which seems to efficiently suppress the non-radiative Auger recombination (see, *e.g.*, Ref. 4 and references therein).

To describe electronic bands in HgTe/CdHgTe QWs, the 4-band Kane model with an axial symmetry (along the growth axis) is traditionally employed. This model proved itself to describe adequately results of magnetotransport [5-8] and magnetooptical [9-19] experiments in *n*-type samples, where effects of strong spin-orbit interaction (for instance, giant Rashba-type spin splitting [8]) were demonstrated. However, experiments performed on *p*-type HgTe/CdHgTe QWs [20-22] have revealed effects that cannot be explained within the axial approximation. In Refs. 21 and 22, these effects of large spin-splitting of electronic states in the valence band were attributed to the symmetry lowering, which emerges due to the anisotropy of the chemical bonds at HgTe/CdHgTe heterointerfaces and which gives rise to a strong mixing of electronic states [23].

Let us also note that the role of BIA and "cubic" terms have been in the past extensively explored and discussed in the context of bulk HgCdTe [24,25] and many other zinc-blende semiconductors [26,27]. However, their impact on electronic states in 2D systems remains much less explored, in particular, when the valence bands in HgTe QWs are concerned. In general, there are at least three reasons while the axial model could be ineffective for the describing some peculiarities observed in magnetotransport and magneto-optical experiments in real HgTe/CdHgTe QWs. The first one is the neglecting of "cubic" terms in the Hamiltonian in the axial approximation. Taking into account the cubic symmetry of the Hamiltonian is important, in particular, at the considering effects of hole populating the side maxima in the valence band (see, e.g. [22]). Two others are bulk inversion asymmetry (BIA) and interface inversion asymmetry (IIA) that also reduce the symmetry of the Hamiltonian and result in significant spin splitting (as large as 10 meV) in the energy spectra (see, e.g., Refs. 9, 28-30 and references therein). All these effects lead to the interaction of the states which are orthogonal in the axial approximation (e.g. Landau levels with different indices), especially in the valence band. On the other hand, we have neglected Rashba terms in our approach. This is justified by quantitative estimates presented in the Supplementary Material [URL will be inserted by publisher].

So far, the clearest experimental evidence of the state mixing is the avoided crossing of "zero-mode" Landau levels (LLs) in HgTe/CdHgTe QWs with an inverted ordering of bands, which was reported in Ref. 28 for HgTe/CdHgTe (001)

---

[*]bovkun@ipmras.ru
[†]milan.orlita@lncmi.cnrs.fr

QWs and reproduced in Refs. 9 and 29 for QWs grown on (013) plane. In the axial model, the wave functions of LLs $n = -2$ and $n = 0$ are orthogonal, and for inverted band structure, these LLs cross at the critical magnetic field $B_c$, which corresponds to the phase transition from the 2D topological insulator to the Quantum Hall effect state [12]. Earlier it has been shown that taking into account the cubic terms in the Hamiltonian practically does not result in interacting and splitting of the above $n = -2$ and $n = 0$ LLs in HgTe/CdHgTe(001) QW, see Fig. 2(b) in Ref. 5. In Refs. 29 and 30, BIA was proposed to be responsible for this effect, however, recent atomistic calculations [23] indicate that the asymmetry due to chemical bonds at the heterointerfaces (IIA) [31] prevails over BIA.

In this work, we present a comprehensive magneto-absorption study of p-type HgTe/CdHgTe QWs, providing solid experimental evidence for the lack of the axial symmetry in the band structure of these systems. In particular, the missing axial symmetry impacts the valence band, which is characterized by the density of states that is considerably larger as compared to the conduction band, and consequently, significantly smaller Landau level spacing. Our experimental data are confronted with results of the band structure calculations performed in both, axial 4-band Kane model and its advanced version, which takes account of IIA and BIA effects.

The paper is organized as follows. The experimental details and the sample description are given in Sec. II. The theoretical basis is introduced in Sec. III while details of the theoretical approach are taken out to the Supplementary materials. The main results of this work are presented and discussed in Sec. IV.

## II. EXPERIMENTAL DETAILS

The samples under study were grown by molecular beam epitaxy on semi-insulating GaAs (013) substrates with an ellipsometric control of the layer thickness and composition [32,33]. A thin ZnTe buffer, thick relaxed CdTe buffer, 40-nm lower $Cd_xHg_{1-x}Te$ barrier, HgTe QW, and 40-nm $Cd_xHg_{1-x}Te$ top barrier were grown one by one without any intentional doping. The heterostructure was completed by a 40-nm-thick CdTe capping layer. Native defects (dominantly mercury vacancies) imply p-type conductivity of as-grown samples, with the hole concentration below $10^{11}$ cm$^{-2}$. The QW width was varied in order to achieve samples with normal, nearly gapless and inverted band structure. The corresponding growth and electrical parameters of all explored samples are presented in Table I.

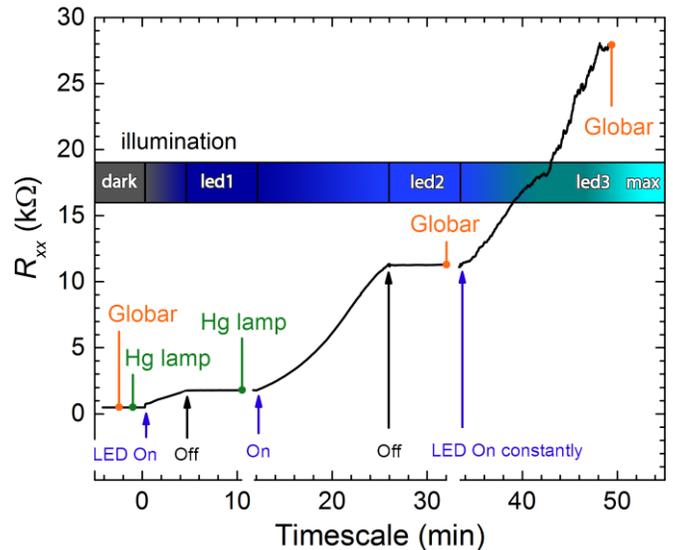

FIG. 1. (Color online) A characteristic dependence of zero-field longitudinal resistance $R_{xx}$ (sample B) during three stages of illumination by a blue LED. Arrows at the bottom indicate times of on/off switching of the blue LED. Vertical lines show times of magneto-optical measurements with an indicated sources and black PE entrance window.

Magneto-optical experiments were performed in the Faraday configuration in magnetic fields up to 11 T delivered by a superconducting coil [9,14,28]. Samples were kept in the low-pressure helium exchange gas at the temperature of 4.2 K. Globar or mercury lamp was used as broadband sources of infrared radiation. The radiation, analyzed by a Fourier transform spectrometer, was guided through black polyethylene (PE) or ZnSe entrance window of the sealed probe, delivered via light-pipe optics to the sample and detected by a composite silicon bolometer placed below the sample. All spectra presented in this paper are relative magneto-transmission, $T_B/T_0$, corrected for the field-induced changes in the response of the bolometer, which is a smooth function of the photon frequency monotonously increasing with $B$.

Transport data were collected using the standard Van der Pauw method simultaneously with the magneto-optical measurements. All the explored samples showed pronounced Shubnikov-de Haas oscillations and the quantum Hall effect in the "dark" state, e.g., without illumination. To change the position of the Fermi energy, the hole concentration in the QW was decreased by means of illumination (taking advantage of the persistent photoconductivity effect, see, e.g., Ref. 13) by a blue light emitting diode (LED) located near the sample. Magneto-optical measurements in the energy range above 80 meV were performed with the ZnSe entrance window on the probe, thus allowing us to illuminate the samples with the middle infrared spectrum of globar (blocked when black PE was used). In the latter case, some of the samples showed nearly insulating behavior with the Fermi level lying within the band gap.

TABLE I. Growth parameters and electrical properties (at $T = 4.2$ K) of the studied samples.

| Sample | QW width, $d$ (nm) | Barrier Cd composition, x | Bandgap, $E_g$ (meV) | Band structure | Hole concentration | |
|---|---|---|---|---|---|---|
| | | | | | without illumination, $p_{max}$ ($10^{10}$ cm$^{-2}$) | under maximal illumination, $p_{min}$ ($10^{10}$ cm$^{-2}$) |
| A (110622) | 4.6 | 75 | 60 | normal | 9.0 | 7.4 |
| B (160126) | 5.0 | 70 | 40 | normal | 7.4 | insulating |
| C (110623) | 5.5 | 62 | 15 | normal | 6.6 | 4.7 |
| D (110624) | 6.0 | 62 | 5 | near gapless | 3.2 | 3.0 |
| E (151214) | 8.0 | 86 | -20 | inverted | 11.0 | 11.0 |

To illustrate this, the evolution of longitudinal resistance of sample B after illumination by a blue LED in three subsequent stages is shown in Fig. 1. In periods without any LED illumination, no changes in the longitudinal resistance were observed, which allowed us to perform magneto-optical and magneto-transport measurements at several fixed hole concentrations. The stages LED$_1$ and LED$_2$ were achieved after two successive doses of illumination (6 and 12 minutes, respectively). The stage LED$_3$ with a rather high value of $R_{xx}$, and therefore with the Fermi level in the midgap position, has been obtained under permanent illumination.

### III. THEORETICAL BASIS

To describe the energy dispersion and Landau level energies in HgTe/CdHgTe QWs, several approaches have been elaborated in the past. In early studies [34], a simple 2-band model with the effective energy gap $E_g^*$ has been used to interpret the observed splitting of the cyclotron resonance (CR) mode. More recently, 8×8 (i.e., 4-band) Kane Hamiltonian [5] has been introduced and successfully employed to describe the energy spectra in both conduction and valence bands with either normal or inverted band ordering [3-9,13-15,28,30,35]. The simplified 2-band model was also explaining gapless and narrow-gap HgTe QWs [36]. Its simplicity appeared in particular convenient for the description of edge states in HgTe QWs with an inverted band structure [36,37]. In this work, we expand the standard 4-band Kane Hamiltonian by including effects of symmetry lowering, which results from the absence of the inversion symmetry in the bulk crystal lattice [29,30,38] as well as from the anisotropy of chemical bonds at HgTe/CdHgTe interfaces [23].

#### A. Hamiltonian and eigenstates

The Landau level spectrum was calculated by the diagonalization of the 8×8 $\boldsymbol{k}\cdot\boldsymbol{p}$ Hamiltonian for (013)-oriented heterostructures, thus considering states with the symmetries $\Gamma_6$, $\Gamma_8$ and $\Gamma_7$ bands [9,13,14]. A tensile strain in individual layers arising due to the mismatch of lattice constants in the CdTe buffer, HgTe QW, and Cd$_x$Hg$_{1-x}$Te barriers were also included, with material parameters taken from Ref. 5. This model allows us to describe the non-parabolic profiles of bands as well as effects of the spin-orbit interaction. The electron and hole states were calculated in two subsequent stages. To this end, the Hamiltonian was represented as a sum of the axial and anisotropic parts:

$$H = H_s + H_a, \quad (1)$$

where $H_s$ is invariant with respect to rotations along the growth axis. In the first stage, the eigenstates of $H_s$ were found [9,13,14]. These eigenstates then served in the second stage as a basis for the expansion of eigenstates of the full Hamiltonian $H$:

$$\psi = \sum_{n,m} c_{n,m} \psi_{n,m}. \quad (2)$$

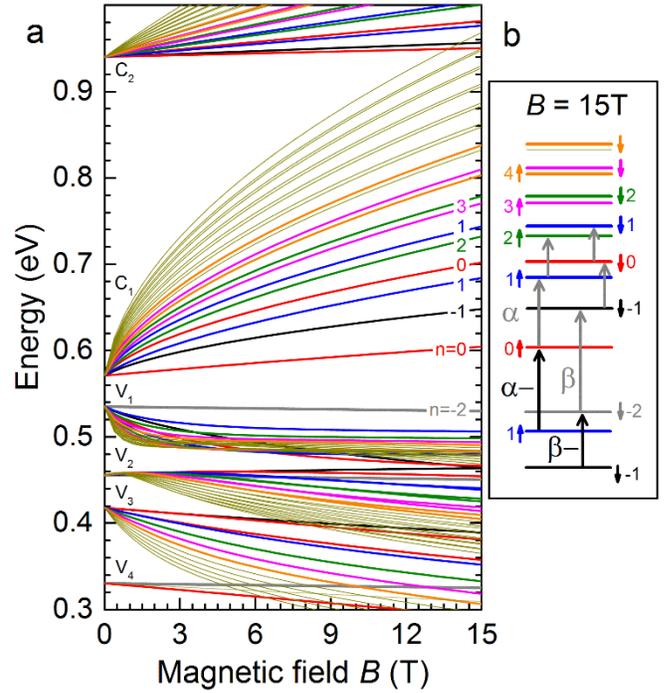

FIG. 2. (Color online) (a) Landau levels for sample B calculated in the axial approximation in two conduction (C$_1$, C$_2$) and four valence (V$_1$ – V$_4$) subbands. (b) Landau level energies in the magnetic field 15 T. Short arrows near the $n$-values left and right from the levels mark the dominant spin orientation. Long arrows indicate allowed electric dipole transitions between LLs with a detailed description in Table II. Transition α (0$_\uparrow$→1$_\uparrow$) are typical for the n-type samples, whilst α– (1$_\uparrow$→0$_\uparrow$), β (-2$_\downarrow$→-1$_\downarrow$) and β– (-1$_\downarrow$→-2$_\downarrow$) are characteristic for the p-type samples.

The corresponding coefficients $c_{n,m}$ were then found using a numerical diagonalization of $H$. Here $n$ denotes the Landau level index and $m$ is the number of the subband. In our calculations, 2 conduction and 11 valence subbands were taken into account, respectively. For each subband, we considered 14 Landau levels ($n$ = –2, –1, 0, 1…11). Further increase in the number of considered subbands and LLs did not provide us with some significant corrections to eigenenergies for magnetic fields exceeding 2 T. At lower fields, the number of LLs taken into account should be increased to achieve a quantitatively correct description of valence states. In particular, this becomes important for HgTe QWs with an inverted band structure (i.e., for widths $d > d_c$) with the maxima of valence subbands typically located at $k \ne 0$.

The anisotropic part of the total Hamiltonian $H_a$ includes terms describing the cubic symmetry, the lack of the inversion symmetry in the bulk crystal and the symmetry lowering at the heterointerfaces. The explicit form of the Hamiltonian in the absence of the magnetic field for the case z // [013] is given in [9], the effect of the magnetic field has been accounted using the Peierls substitution. The BIA term was derived from 14×14 Kane Hamiltonian proposed in Ref. 34, from which also the corresponding parameters were taken (same for CdTe and HgTe). Explicit expressions for the BIA terms may be found in Supplemental Materials [URL will be inserted by publisher]. The IIA terms for the (013)-oriented QWs are expressed in Ref. 22 (see Eqs. 5 and 6 therein). The parameter $g_4$ (solely determining the power of the effect [22]) was taken as 1.4 eV×Å in our advanced Kane model. The specified value gives exactly the same energy of Dirac cone splitting as calculated in the tight-binding approximation [23] in a (001)-oriented HgTe QW with the critical width $d_c = 6.3$ nm at $k = 0$. Indeed, for this symmetric crystallographic orientation taking into account BIA corrections to the Hamiltonian practically does not open the gap at the critical QW width (see, e.g. [30]), so the splitting of the Dirac cone results from the IIA only.

TABLE II. Squared matrix elements of allowed optical transitions between Landau levels in the axial approximation. Results are given for the lowest energy conduction subband ($C_1$) and the highest valence ($V_1$) subband at a magnetic induction of 6T for the normal (Sample B) and inverted (Sample E) band structure.

| Landau level | | | Transition label | $\Delta n$ | $|r_{f,i}|^2$ (Å$^2$) | |
|---|---|---|---|---|---|---|
| initial | | final | | | Sample B | Sample E |
| $C_1$ | $0\downarrow$ | $C_1$ $1\downarrow$ | $\varepsilon$ | 1 | 10385 | 11082 |
|  | $1\uparrow$ |  $2\uparrow$ | $\delta$ | 1 | 8287 | 7720 |
|  | $-1\downarrow$ | $0\downarrow$ | $\gamma$ | 1 | 5724 | 6665 |
|  | $0\uparrow$ | $1\uparrow$ | $\alpha$ | 1 | 3766 | 3184 |
| $V_1$ | $-1\downarrow$ | $V_1$ $-2\downarrow$ | $\beta-$ | $-1$ | 2945 | 1653 |
|  | $0\downarrow$ | $-1\downarrow$ | $\gamma-$ | $-1$ | 5722 | 5670 |
|  | $2\uparrow$ | $1\uparrow$ | $\delta-$ | $-1$ | 5388 | 1319 |
|  | $3\uparrow$ | $2\uparrow$ |  | $-1$ | 8476 | 8337 |
| $V_1$ | $-2\downarrow$ | $-1\downarrow$ | $\beta$ | 1 | 1366 | 2576 |
|  | $1\uparrow$ | $0\uparrow$ | $\alpha-$ | $-1$ | 1109 | 2566 |
|  | $2\uparrow$ | $C_1$ $1\uparrow$ |  | $-1$ | 247 | 200 |
|  | $0\downarrow$ | $-1\downarrow$ |  | $-1$ | 219 | 530 |
|  | $-1\downarrow$ | $0\downarrow$ |  | 1 | 186 | 42 |

### B. Matrix elements for optical transitions between Landau levels

Matrix elements for optical transitions between Landau levels were calculated for unpolarized radiation. The probability of such a transition is proportional to the square of the dipole moment matrix element (product of the electron charge and the matrix element of the coordinate operator). The matrix element $r_{f,i}$ of the coordinate operator between the initial $|i\rangle$ and final $|f\rangle$ states satisfies the equation:

$$\boldsymbol{v}_{f,i} = \frac{i}{\hbar}[H,\boldsymbol{r}]_{f,i} = \frac{i}{\hbar}(\varepsilon_f - \varepsilon_i)\boldsymbol{r}_{f,i}, \qquad (3)$$

where $\boldsymbol{v}$ is the velocity operator. First, one has to find the velocity operator using Eq. 1, and then, calculate its matrix elements, see the Supplemental Materials at [URL will be inserted by publisher] for the explicit form of the velocity operator.

In the axial model, the selection rules for electric-dipole excitations are $\Delta n = \pm 1$ and no spin-flip excitations are allowed. The applicability of these "axial" selection rules for HgTe/CdHgTe QWs was demonstrated in a number of works [9,12-15,28]. It is worth mentioning that the interpretation of (cyclotron resonance-like) excitations within the conduction band is simpler due to a rather large spacing of LLs. In contrast, the identification of particular cyclotron resonance excitations in *p*-type samples is more challenging. The relatively flat valence bands imply a rather high density of states, and when the magnetic field is applied, the rather dense spacing of LLs, and consequently, a number of possibly contributing excitations. To facilitate this identification, we calculated the corresponding matrix elements for electric-dipole transitions between different pairs of considered LLs to be able to compare the relative strength of excitations.

Table III. Calculated squared matrix elements $|r_{f,i}|^2$ for dominant optical transitions in sample B versus the magnetic induction (in units of Å$^2$).

| B (T) | 1 | 2 | 3 | 4 | 5 | 6 | 7 | 8 | 9 | 10 |
|---|---|---|---|---|---|---|---|---|---|---|
| $\beta-$ | 21981 | 10127 | 6418 | 4644 | 3614 | 2945 | 2478 | 2133 | 1869 | 1660 |
| $\beta$ | 4276 | 2964 | 2284 | 1863 | 1576 | 1366 | 1207 | 1081 | 979 | 894 |
| $\alpha-$ | 3818 | 2534 | 1907 | 1533 | 1286 | 1109 | 977 | 875 | 793 | 727 |

We have considered optical transitions among 30 different Landau levels: 4 low-energy LLs in $C_1$ subband, 11 top LLs both in $V_1$ and $V_2$ subbands ($n = -2, -1, 0, .., 4$) and 4 LLs in $V_3$ subband ($n = -1, 0$), see Fig. 2. Excitations from/to LLs in other distant subbands were not included since they do not fit into the spectral window explored in our experiments. The transitions originally "forbidden" in the axial model may become active due to mixing neighboring Landau levels induced by BIA or IIA effects. Since these symmetry-lowering effects profoundly modify the profiles of valence subbands, one may expect that they primarily influence excitations involving LLs originating from those valence subbands.

### IV. RESULTS AND DISCUSSION

Before we start a detailed discussion of the experimental data, let us summarize the main trends/conclusions found in the theoretical calculations. In Table II, we show the squared matrix elements of the coordinate operator $|r_{f,i}|^2$ calculated at $B = 6$ T for QWs with a normal and inverted band ordering. Obtained values allow us to estimate the mutual intensity of individual absorption lines. In addition, we also provide the square of the coordination matrix elements for $\beta$, $\beta-$ and $\alpha-$ lines for selected values of $B$ (Tab. III). This latter table shows that theoretically, no significant changes in mutual intensities are expected with the magnetic field. Let us also note that the squared coordination matrix element has to be multiplied by the LL degeneracy (linear in B) and the transition energy to get the field-dependence of the total oscillator strength for a given excitation.

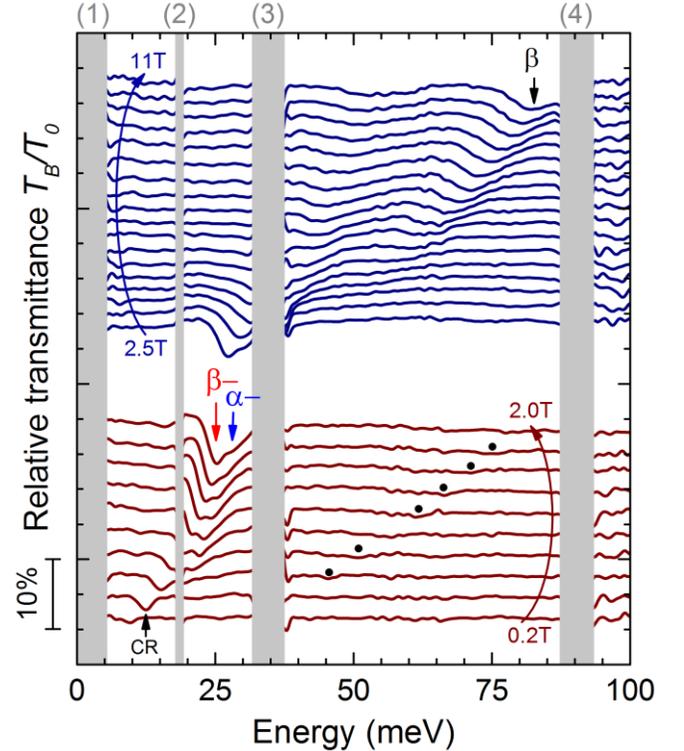

FIG. 3. (Color online) Low-temperature magneto- transmission spectra collected on sample D. The Hg lamp was used as the source of the radiation for the spectra measured at the fields below 2 T (with the step of 0.2T). At higher fields, the globar was employed and the step was increased to 0.5 T. Dominant optical transitions are denoted by vertical arrows and labeled according to Table II. The relatively broad spectral feature marked by black dots correspond to a series of excitations analogous to η lines observed and analyzed in detail for the sample B (cf. Fig. 8 and Tab. IV). The grey areas correspond to the spectral regions fully opaque due to the reststrahlen band of HgCdTe and GaAs (2 and 3, respectively) and absorption in the multilayer beamsplitter (4). In the low-energy region (1), the used beamsplitter is transparent but rather inefficient for Fourier-transform experiments.

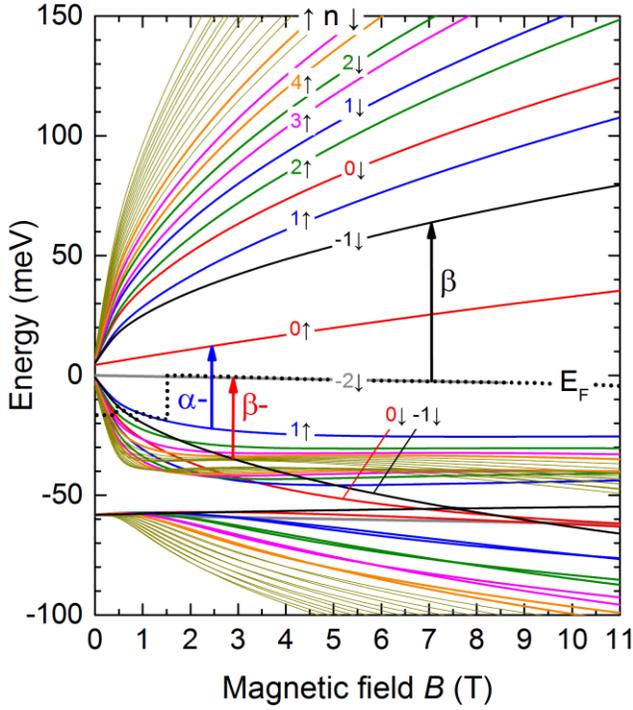

FIG. 4. (Color online) LLs calculated within the axial model. Dominant optical transitions are denoted by vertical arrows.

One may immediately see in Table II that the strength of intraband (cyclotron resonance-like) transitions ($C_1 \to C_1$ and $V_1 \to V_1$) exceeds that of interband transitions ($V_1 \to C_1$). Besides, the energies of the intraband transitions are less sensitive, as compared with interband excitations, to the QW width and to fluctuations of $E_g$ resulting from the lateral inhomogeneity of heterostructures. This implies that the dominant contribution to the magneto-optical response should stem from intraband excitations, which should also give rise to narrower spectral lines.

Let us also note that electric-dipole transitions within the given electron or hole subband (within $C_{1,2}$ or within $V_{1,2}$) always follow the selection rules $\Delta n = 1$ and $-1$, respectively. Such transitions represent pure cyclotron resonance modes and these selection rules reflect the opposite sense of the cyclotron motion of electrons and holes, and consequently, also the opposite circular polarization of absorbed radiation. In the inter(sub)band response, excitations may be active in both circular polarizations: $\Delta n = \pm 1$. For example, transitions to LL $n = -1_\downarrow$ in the $C_1$ subband are allowed from both the top LL $n = -2_\downarrow$ in the $V_1$ subband and from deeper LL $n = 0_\downarrow$ (see Table II). In the latter case, the squared matrix element $|r_{f,i}|^2$ is significantly weaker mainly due to bigger energy difference ($\varepsilon_f - \varepsilon_i$) for the given transition (see Eq. 3).

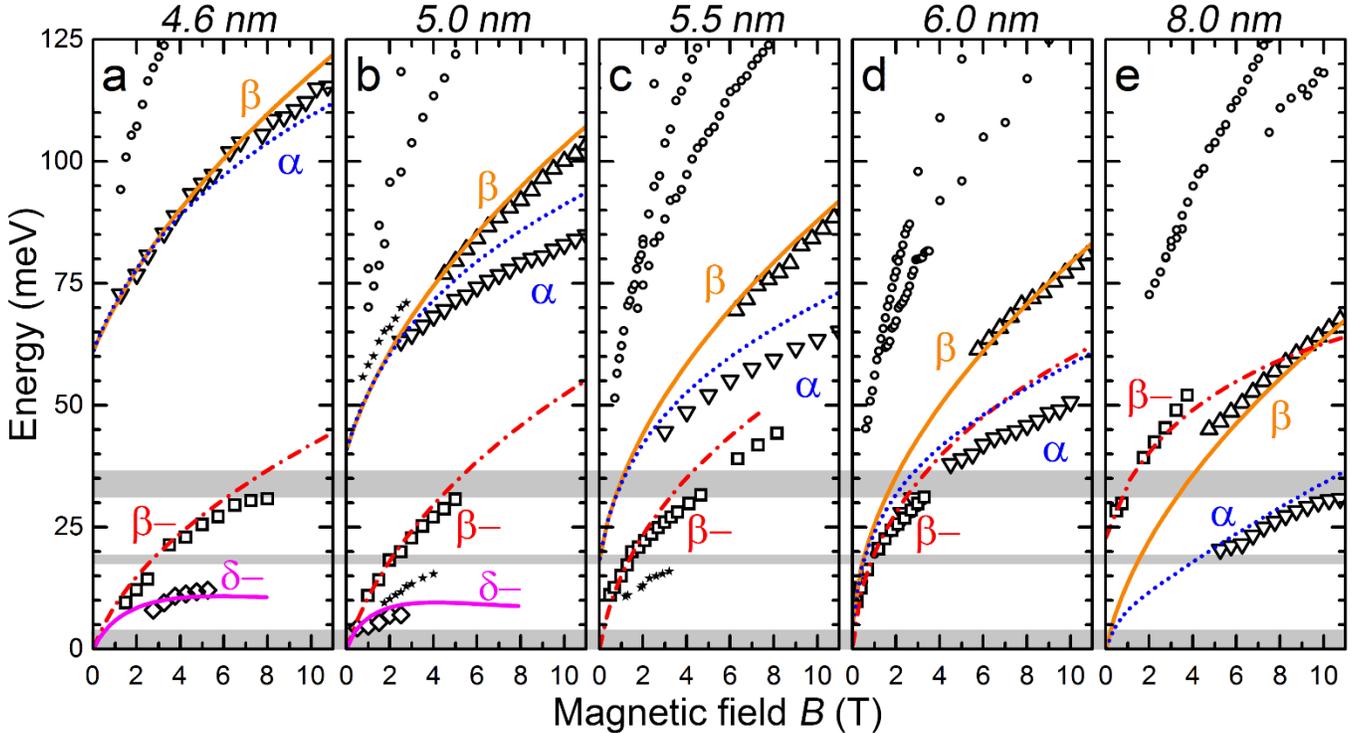

FIG. 5. (Color online) Energy of α (blue dotted), β (orange solid), β– (red dash-dot) and δ– (magenta solid) transitions as a function of magnetic field of HgTe/CdHgTe QWs with normal a) $p_s = 9.0 \times 10^{10}$ cm$^{-2}$, b) $p_s = 7.4 \times 10^{10}$ cm$^{-2}$, c) $p_s = 6.6 \times 10^{10}$ cm$^{-2}$, d) $p_s = 3.2 \times 10^{10}$ cm$^{-2}$, and inverted e) $p_s = 1.1 \times 10^{11}$ cm$^{-2}$ band structure ($d_{qw}$ is shown on top of each graph). The experimental data are represented by symbols: down triangles, up triangles, squares and diamonds for the α, β, β–, δ– transitions. Circles represent high-energy transitions discussed below; stars represent additional transitions due to the reduced symmetry of the system (discussed later as well).

## A. Dominant optical transitions

We will now proceed with the interpretation of our experimental data in two subsequent steps. In the first one, we analyze the data using the axial model, thus identifying limits of such an approach. In the second step, we compare experiments with a more sophisticated model, which includes both above-mentioned mechanisms of symmetry lowering.

The magneto-transmission spectra recorded on sample D are plotted in Fig. 3. The "dark" hole concentration in this sample is $3.2 \times 10^{10}$ cm$^{-2}$ that corresponds to the LL filling factor $\nu \approx 1$ at $B = 1.3$ T. LLs calculated in the axial approximation are plotted in Fig. 4, where the dotted line indicates the expected position of the Fermi energy as a function of $B$. The transition, which gradually emerges in the spectrum and becomes dominant at $B \approx 1$ T, may be identified as the β line, see Fig. 4. It corresponds to the excitation of electrons from the topmost (partially populated) valence LL $n = -2_\downarrow$, to $n = -1_\downarrow$ level in the $C_1$ conduction subband. This transition is characteristic of all $p$-doped HgTe/CdHgTe QWs and has been identified already in the early work dedicated to magneto-optical properties of such systems [12]. The response at low energies and low magnetic fields is dominated by α– and β– transitions. The former line is, in this case, the lowest-in-energy interband excitation, from $n = 1_\uparrow$ LL to the zero-mode $n = 0_\uparrow$ level. This line is analogous to the α line in the conduction band of QWs with an inverted band structure [9,13-15,28,29,35]. The β– line represents a purely cyclotron resonance mode, $-1_\downarrow \rightarrow -2_\downarrow$, which may be expected in all $p$-doped samples (for a particular hole concentration).

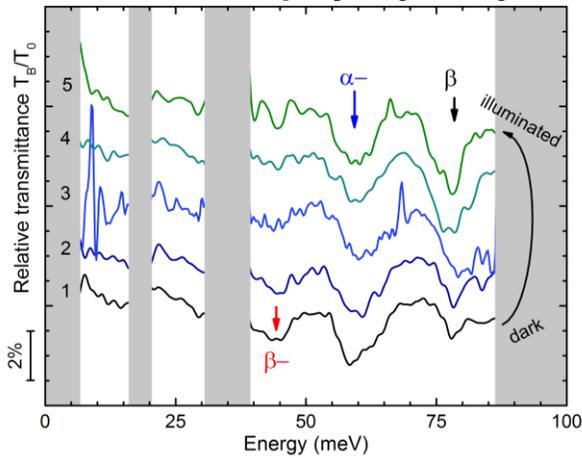

FIG. 6. (Color online) Transmission spectra of sample C in the magnetic field $B = 8$ T measured at various hole concentrations from $p_{max}$ (1) to $p_{min}$ (5) (spectra are shifted with the step of 2% for clarity).

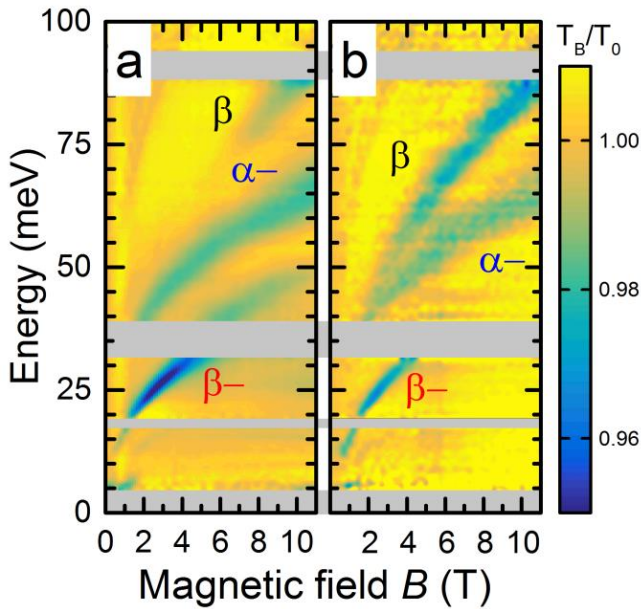

FIG. 7. (Color online) Magnetoabsorption of sample C for hole concentrations $p_{max}$ (a) and $p_{min}$ (b) plotted as false color-maps. Dominant absorption labeled as β–, α– and β.

The β, α– and β– lines dominate the magneto-optical response of all investigated samples (see Fig. 5). In case of a normal band structure, transitions α– and β have an interband character and may thus be always observed in magnetic fields high enough, when the quantum limit of our (weakly $p$-doped) samples is approached. The zero-field extrapolation of their positions provides us with a good estimate of the bandgap. In agreement with theoretical expectations, the band gap $E_g$ indeed decreases with the thickness of QWs with the normal band ordering as shown in Fig. 5(a-d). In contrast to these interband excitations, the β– line has a purely intraband (CR-like) character, with the strength proportional to the total carrier (hole) density in QWs.

The correct assignment of the dominant absorption lines may be independently checked in the spectra recorded after various illumination using the blue diode (due to PPC effect), which gradually lowers the total hole density. The relative magneto-transmission spectra plotted in Fig. 6 show the evolution of the line intensities after several illumination steps. One may immediately notice that the β– line gradually disappears from the spectrum under illumination, following thus the decreasing hole density. The interband excitations α– and β show distinctively different behavior. The intensity of α– transition remains constant with illumination and the strength of the β transition even increases.

The observed variation of intensities with the illumination may be explained when the specific $n = -2_\downarrow$ zero-mode LL is considered, see Fig. 4. This level represents the final and initial states for β– and β excitations, respectively, but at the same time, it is not involved in the α– transition. In the quantum limit, when the only $n = -2_\downarrow$ level is occupied by holes (*e.g.*, at $B = 8$T for sample C, see Fig. 6), the illumination by the blue LED decreases this occupation, or equivalently, it increases the number of electrons in this level. This directly implies a gradual increase and decrease of the absorption strengths for the β and β– excitations, respectively. The α– transition, with the initial and final states in the $n = 1_\uparrow$ and $n = 0_\uparrow$ LLs of the $V_1$ and $C_1$ subbands, respectively, remains unaffected.

The above-discussed interplay between intensities of β– and β lines can be traced in a wide range of magnetic fields. At a higher hole concentration, see Fig. 7(a), the β– line is observed up to 11 T, while the β line only appears at magnetic fields above 6 T, reflecting thus the gradually increasing number of electrons in $n = -2_\downarrow$ LL. After illumination, *i.e.*, with the hole density lowered, the β line emerges in the spectrum at magnetic fields as low as $B \approx 3$ T.

To complete our discussion, let us note that the intensity of the β– line decreases above 6 T. As a matter of fact, the β–line represents the cyclotron resonance absorption in the quantum limit of the explored QWs (with only the highest hole LL occupied). In such a case, the intensity of the line is proportional to the hole density (presumably constant with B), the square of the coordination matrix element (expected to roughly follow 1/B dependence) and the transition energy (increasing sub-linearly with B), which implies a weak decrease of the β– line intensity. This can be also shown in numbers [see Tab. III and Fig. 7(b)]: the increase of the transition energy of 30, 43 and 54 meV at B = 3, 6 and 9 T does not compensate the decrease in the square of the coordination matrix $|r_{f,i}|^2 \approx 6400/3000/1900$ (Å$^2$). Apart from the decrease in intensity we address broadening of the absorption line (the $B^{1/2}$ dependence of the widths is expected for LLs with a Gaussian profile [35,36]) and to LL mixing in the valence band due to the symmetry lowering effects (see below). The decrease of the β– line strength was observed for all QWs [Fig. 5(a-d)]. In contrast, the intensity of the β line seems to be increasing in its strength.

### B. High-energy interband transitions

The deeper analysis of the transmission spectra of sample B at higher photon energies ($\hbar\omega > 100$ meV) shows the presence of at least seven additional transitions, which correspond to excitations between LLs in $V_1$ and $C_1$ subbands and of two pronounced excitations, which are identified as transitions between pairs of LLs in the $V_3$ and $C_1$ subbands (see Fig. 8). The extrapolation of the line positions to zero magnetic field yields the separation of subbands of about 40 meV and 160 meV, respectively, which is in good agreement with the calculations performed within the axial model. As for transitions from $V_2$ to $C_1$ subbands, they are supposed to be relatively weak due to opposite parity of corresponding wave functions (strictly speaking valid for $k=0$ parity only). A weak absorption line may be traced in the spectra at magnetic fields high enough [above 5 T, see Fig. 8(c)], which can be associated with the α+ transition, see Table IV.

The relative intensities of high-energy absorption lines do not exceed 2%. The signal-to-noise ratio in our experimental data allowed us to reveal absorption lines with relative intensities down to 0.5%. It is important to note that our calculations predict fairly low intensities for these high-energy transitions, with the squared matrix elements reduced by almost two orders of magnitude as compared to dominant transitions (cf. data in Tables II&IV). Nevertheless, these transitions are still visible in the spectra, since they are rather closely spaced, and in fact, strongly overlapping in our experiments (taking account of the experimentally observed line widths). For example, the observed line $\eta_1$ in Fig. 8(c) results from two different transitions with nearly identical energies as shown in Fig. 8(e).

As for the interband transitions from the $V_3$ subband, we attribute the lower line $\tau_1$ to the transition $-1_\uparrow \rightarrow 0_\uparrow$, and $\tau_2$ to the transition $0_\downarrow \rightarrow -1_\downarrow$. In this case, let us note that the energies of initial-state LLs in the $V_3$ subband depend on the magnetic field only very weakly, in contrast to the final-state LLs in the conduction band. This way, the field-dependence and separation of LLs in the $C_1$ conduction band are thus straightforwardly visualized. This gives us a possibility to study conduction-band LLs even in *p*-type samples, which is an alternative approach to the standard cyclotron resonance technique applied to *n*-doped QWs [9,11,13-15,29,34,35].

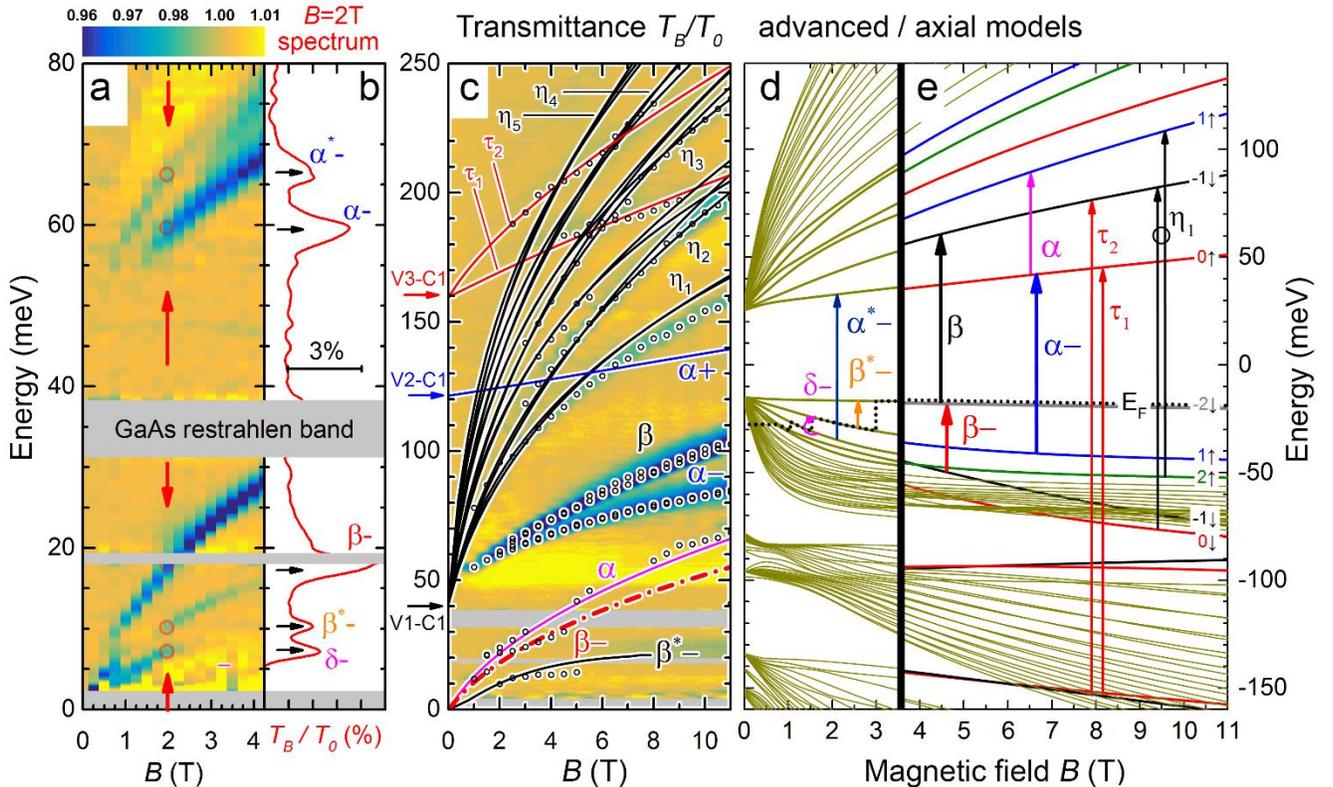

FIG. 8. (Color online) Magnetoabsorption of sample B plotted as false-color plots for the maximal (a) and minimal (c) hole concentrations. Red vertical arrows and circles in (a) show the transitions clearly resolved and denoted by Greek letters in the magneto-absorption spectrum taken at B = 2T and plotted in (b). The solid lines in (c) correspond to the calculated energies of transitions β*–, β–, α, α+, $\tau_1$, $\tau_2$ and absorption lines $\eta_1$, $\eta_2$, $\eta_3$, $\eta_4$, see the main text. Black open circles stand for the observed transmission minima. (d,e) The LL fan charts calculated in the axial (e) model for B > 3.5T, where all indicated transitions follow the standard selection rules $n \rightarrow n \pm 1$ and the advanced model (d) for B<3.5T, where LLs are mixed allowing for additional transitions in the electric dipole approximation. The dotted line indicates the Fermi level position of the "dark" state (maximal hole concentration). The arrows and Greek letters denote LL transitions, observed in the magnetoabsorption spectra, absorption line $\eta_1$ contains two close in energy transitions $0_\downarrow \rightarrow -1_\downarrow$ and $-2_\uparrow \rightarrow 1_\uparrow$ connected with a small circle.

Let us conclude this subsection by the observation that the band structure model preserving axial symmetry provides us with an overall acceptable, but as discussed later on not with a complete quantitative description of interband excitations between LLs in different valence and conduction subbands. In the following part, we will specifically focus on limits of the axial model.

### C. Effects of symmetry lowering in low-fields

In the previous part, we have shown that the axial model is capable of explaining the dominant features in the magneto-absorption spectra of HgTe/CdHgTe QWs. In what follows, we will concentrate on the findings that cannot be explained within the axial approximation.

Let us start with the magneto-optical response at relatively low energies, which is dominated by the cyclotron resonance of charge carriers (holes) present in our samples. According to the calculations performed within the axial 4-band Kane model, such a response should be at relatively low hole densities (for the filling factor $\nu < 2$) dominated by two CR modes, by the $\beta-$ and $\delta-$ lines. These transitions are electric-dipole-allowed within the axial model and correspond to inter-LL excitations, $-1_\downarrow \to -2_\downarrow$ and $2_\uparrow \to 1_\uparrow$, respectively [Fig. 8(d,e)]. Indeed, both these modes are clearly resolved in our experimental data, with the mutual relative intensity depending on the applied magnetic field and particular hole density [Fig. 6(a,b) and Fig. 8(a,b,c)].

Table IV. Squared matrix elements of allowed optical transitions between Landau levels for subbands $V_1$, $V_3$ and $C_1$ in the axial approximation at a magnetic induction of 6T for sample B.

| Landau level | | Transition label | $\Delta n$ | $|r_{f,i}|^2$ ($e$ Å$^2$) | $\Sigma |r_{f,i}|^2$ |
|---|---|---|---|---|---|
| initial | final | | | | |
| $V_1$  $2_\uparrow$  $C_1$  $1_\uparrow$ | | $\eta_1$ | $-1$ | 493 | 930 |
|   $0_\downarrow$   $-1_\downarrow$ | | | $-1$ | 437 | |
| $V_1$  $-1_\downarrow$  $C_1$  $0_\downarrow$ | | $\eta_2$ | 1 | 372 | 1241 |
|   $1_\uparrow$   $0_\uparrow$ | | | $-1$ | 341 | |
|   $3_\uparrow$   $2_\uparrow$ | | | $-1$ | 263 | |
|   $1_\uparrow$   $2_\uparrow$ | | | 1 | 265 | |
| $V_1$  $2_\downarrow$  $C_1$  $1_\downarrow$ | | $\eta_3$ | $-1$ | 258 | 779 |
|   $0_\downarrow$   $1_\downarrow$ | | | 1 | 202 | |
|   $2_\uparrow$   $3_\uparrow$ | | | 1 | 227 | |
|   $4_\uparrow$   $3_\uparrow$ | | | $-1$ | 92 | |
| $V_1$  $3_\downarrow$  $C_1$  $2_\downarrow$ | | $\eta_4$ | $-1$ | 213 | 417 |
|   $3_\uparrow$   $4_\uparrow$ | | | 1 | 75 | |
|   $1_\downarrow$   $2_\downarrow$ | | | 1 | 129 | |
| $V_1$  $4_\downarrow$  $C_1$  $3_\downarrow$ | | $\eta_5$ | $-1$ | 77 | 199 |
|   $4_\uparrow$   $5_\uparrow$ | | | 1 | 67 | |
|   $2_\downarrow$   $3_\downarrow$ | | | $-1$ | 55 | |
| $V_2$  $-1_\uparrow$  $C_1$  $0_\uparrow$ | | $\alpha+$ | $-1$ | 368 | – |
| $V_3$  $-1_\uparrow$  $C_1$  $0_\uparrow$ | | $\tau_1$ | 1 | 343 | – |
| $V_3$  $0_\downarrow$  $C_1$  $-1_\downarrow$ | | $\tau_2$ | $-1$ | 128 | – |

A closer inspection of the data, however, reveals another CR mode denoted $\beta^{*}-$ in Fig. 8(a,b,c), located in between the $\beta-$ and $\delta-$ lines and characterized by roughly half the intensity. This line is observed in the response of several samples with a normal ordering of bands, and it is best manifested for sample B, which shows the highest mobility from the series of samples explored. Importantly, this line cannot be, having in mind the occupation of valence band LLs with relatively low-indices, explained within the axial 4-band Kane model. At the same time, its spectral position fairly well matches the transition $1_\downarrow \to -2_\downarrow$, see Fig. 8(d), which is strictly forbidden within the axial model.

The situation changes when symmetry lowering effects are included in the corresponding Hamiltonian. The resulting band structure then loses its axial symmetry. Among other effects, the original selection rules $\Delta n = \pm 1$ is broken for inter-LL excitations and the magneto-optical response may become significantly richer concerning the number of electric-dipole allowed transitions. Alternatively, the impact of symmetry lowering effects may be also understood in terms of mixing of the original (in the axial model calculated) Landau levels. This mixing becomes in particular important in valence subbands, where the large density of states implies rather narrow spacing of Landau levels.

To demonstrate this, we have estimated intensities of CR-like excitations into the highest in energy valence Landau level ($n = -2_\downarrow$ in the axial model) from several subjacent LLs, denoted as $v_1...v_6$ in Fig. 9. Due to the LL mixing induced by the symmetry lowering effects, the strength of the $-1_\downarrow \to -2_\downarrow$ excitation, which is the only one allowed within the axial model, is now distributed among a series of excitations from different initial LLs to $n = -2_\downarrow$ level. The relative strength of these excitations strongly varies with *B*. Importantly, the originally electric-dipole forbidden transition $1_\downarrow \to -2_\downarrow$ [line 1 in Fig. 9] becomes active in a rather broad range of fields, with the transition probability smaller but still comparable to the strength of the $-1_\downarrow \to -2_\downarrow$ excitation (*e.g.*, equal to 1/3 at 2 T). We may also notice that the maximum of the oscillator strength always corresponds to the carrier excitation from the initial level that is closest to the $n = -1_\downarrow$ LL. This way, the response approaches with increasing $B$, and consequently, with the enhanced spacing of LLs and their suppressed mixing, the expectations of the axial model, in which the β– line dominates the low-energy response.

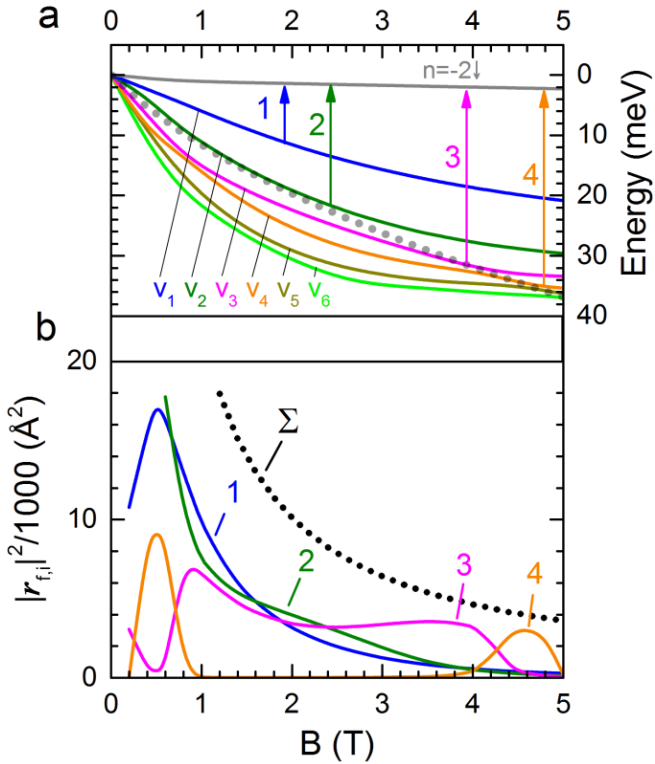

FIG. 9. (Color online) (a) Landau levels for sample B calculated in the advanced model with symmetry lowering effects, "axial" level $-1_\downarrow$ is depicted with the dotted line. Arrows represent transitions discussed in the text. (b) Squared matrix elements versus the magnetic field for transitions from "mixed" LLs $v_1$-$v_6$ to the top valence band LL $-2_\downarrow$, shown in Fig. 9(a). Dotted line $\Sigma$ stands for "axial" $|r_{f,i}|^2$ of the $\beta-$ transition.

The appearance of the weak $\alpha^*$– line in the magneto-absorption spectrum serves as another experimental evidence for symmetry lowering effects [Fig. 8(a,b,d)]. The positions of this line match well the spin-flip transition $-1_\downarrow \to 0_\uparrow$, which is clearly electric-dipole forbidden in the axial approximation. Nevertheless, its presence may be again explained by mixing of LLs due to symmetry lowering effects. In this case, it is the mixing of $n = 1_\uparrow$ and $n = -1_\downarrow$ levels, which makes this excitation electric-dipole active. This line may be viewed as a satellite transition of the α– line. Notably, the $\alpha^*$– line gradually disappears from the spectra with the increasing magnetic field (completely above 4T), when the spacing of $n = 1_\uparrow$ and $n = -1_\downarrow$ levels increases and the mixing effect thus weakens. Let us also note that the $\alpha^*$– line is also present in the response at low hole concentrations. Nevertheless, it is not directly visible in the false-color plot in Fig. 8(c). This is due to a pronounced high-energy tail of the β line, the intensity of which is greatly enhanced as compared to Fig. 8(a,b) because of the increased population of LL $n = -2_\downarrow$ (by electrons).

### D. Effects of "zero-mode" LL avoided crossing

The symmetry lowering effects are also clearly manifested by an avoided crossing of so-called zero-more LLs: $n = -2_\downarrow$ and $n = 0_\uparrow$. Nevertheless, their impact on magneto-optical response has been so far studied only in *n*-type samples. The very first observation of this avoided crossing was reported on HgTe/CdHgTe (001) QWs [28]. The splitting was tentatively attributed to BIA, nevertheless, a possible influence of electron-electron (*e-e*) interaction was not excluded. Later on [29], an analogous effect has been observed in (013)-oriented samples and also assigned to BIA. Nevertheless, to achieve quantitative agreement with experimental data, the term describing BIA in the corresponding Hamiltonian had to be taken 3× larger as compared to the values known for CdTe. The avoided crossing of zero-mode LLs was clearly visible via α' and β' absorption lines [Fig. 11(c)], for which these zero-mode LLs represent initial states.

Here we report the avoided crossing of zero-mode LLs in *p*-doped samples, see Fig. 10 for the magneto-absorption data collected on sample E. Just the observation of the avoided crossing with the magnitude comparable to values reported in Refs. 24 and 25 for samples with the opposite (and higher) doping make *e-e* effects unlikely to explain its appearance.

Similar to samples with the normal ordering of bands, the magneto-optical response of sample E with an inverted band gap is dominated by three absorption lines. These are attributed (as justified *a posteriori*) to α–, β and β– lines, see Fig. 10. Notably, in contrast to samples with a normal band ordering, the β– transitions become an interband excitation, while α– and β transitions have an intraband character. The observed response changes rather dramatically its character around the field of 5T, when the intensity of the β– excitation drops relatively fast and it is replaced by α– and β transitions. This relatively sudden change in the response correlates well with our theoretical expectations. Indeed, the (avoided) crossing of zero-mode LLs is expected at the critical field of $B_c \approx 4.8$ T, as shown in the LL spectrum calculated the "advanced" model, see Fig. 11(c).

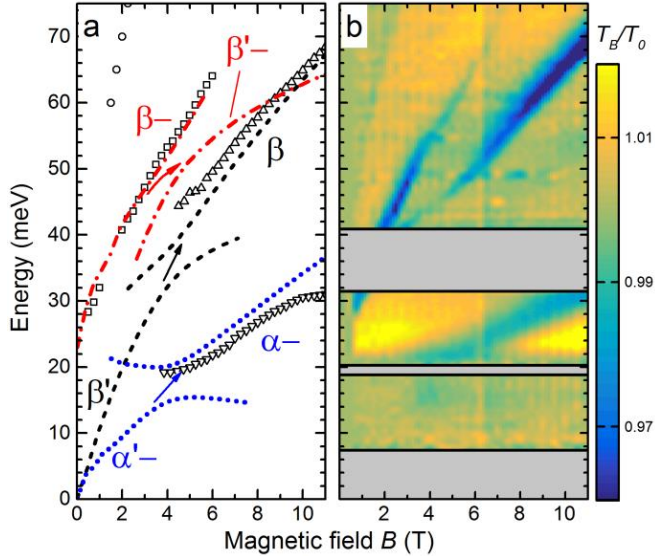

FIG. 10. (Color online) (a) The energies of α, β and β– transitions as a function of magnetic field. The arrows represent transition energy for the case of uncoupled LLs. We show "tails" of avoided crossing only in the region from 2 to 7T, since the "mixing" effects are negligible furthermore. The experimental data are represented by down triangles, up triangles and squares for α, β and β–, respectively. (b) Magnetoabsorption of sample E plotted as a false-color plot. The opaque regions masked using horizontal grey areas.

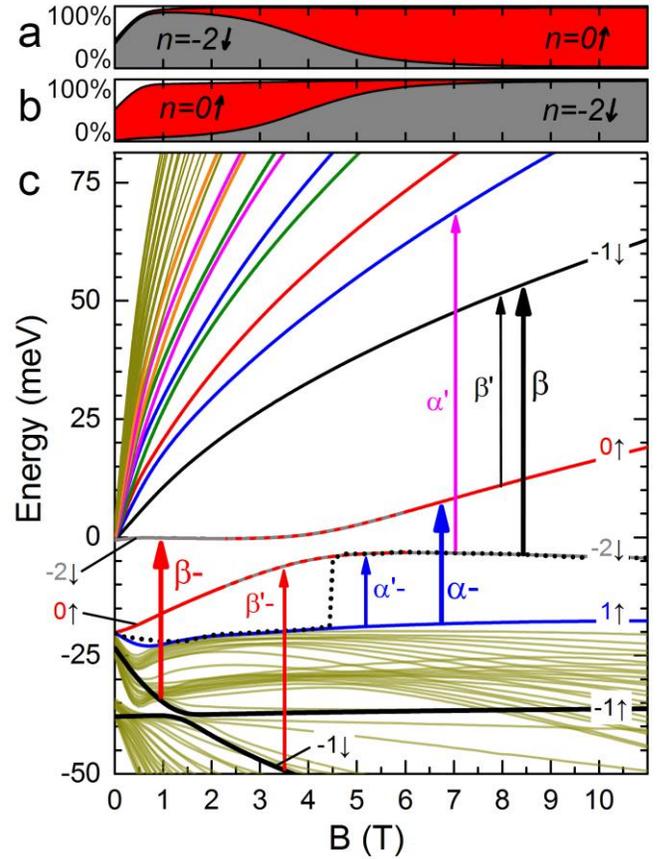

FIG. 11. (Color online) (a,b) The admixture of $n = -2_\downarrow$ (gray) and $n = 0_\uparrow$ (red) states from the axial model in the upper (a) and lower (b) zero-mode Landau level, anticrossed due to the symmetry lowering effects. (c) Landau levels calculated within the advanced model. The arrows and Greek letters denote LL transitions: bold for observed in the magnetoabsorption spectra, while thin is not observed but discussed in the text. Fermi level is shown with dotted line. Particular Landau levels are marked with $n$ and dominant spin orientation.

Let us now focus on important details of the response, which is typical of p-type samples (as compared to n-type specimens in Refs. 28 and 29). According to the magneto-transport characterization of sample E (ν=1 at B ≈ 4.5T), the hole density (without any illumination) is rather low, close to $1.1 \times 10^{11}$ cm$^{-2}$, which implies an approximate field-dependence of the Fermi level indicated by the dotted line in Fig. 11(c).

Similar to samples with normal band ordering (see Fig. 5), the line dominating the response at low magnetic fields is indeed the β– excitation, *i.e.*, $-1_\downarrow \rightarrow -2_\downarrow$ transition in the axial model. Moreover, its position matches pretty well the theoretically calculated one (Fig. 10(a)). When the critical field $B_c$ is approached, the β– transition deviates from axial calculations and weakens rather fast, which can be attributed to the symmetry lowering effects only. To characterize the mixing quantitatively, we have calculated, and plotted in Fig. 11 (a,b) the admixture of $n = -2_\downarrow$ and $n = 0_\uparrow$ states from the axial model in the anticrossed zero-model LLs. One can immediately see that the lower zero-mode LLs is dominantly composed of $n = 0_\uparrow$ state at low magnetic fields. With increasing B, the admixture of $n = -2_\downarrow$ (gray) state growth, reaching 90% at B = 6 T. The intensity of the β– line is proportional to the admixture of the $n = -2_\downarrow$ state in the upper zero-mode LL (the lowest level of the conduction band). As clear from Fig. 11(a), $n = -2_\downarrow$ state indeed dominates (around 90%) in this upper zero-mode LLs. However, with the increasing B, it becomes replaced fast by $n = 0_\uparrow$ state, thus representing only 10% at B = 6 T. Hence, the β– transition diminish fast from the magneto-transmission spectrum, contrary to expectations based on the axial model. At the same time, the lower zero-mode LL becomes almost fully occupied around $B = B_c$, which does not allow β'– transition to appear in the response.

Above the critical field of $B_c$, two strong resonances appear in the response. Following our calculations, the upper one may be identified as the β transition. It becomes active due to (partial) occupation of $n = -2_\downarrow$ level at $B > B_c$. The line at lower energies matches well with the α– transition as shown in Fig. 10(a). Also for β and α– transitions, one may trace a rather clear impact of mixing effects in spectra. These are best visible around the critical field $B_c$, when the field dependence flattens significantly. Without the symmetry lowering effects, and consequently, without the avoided crossing of zero-mode LLs, such a flattening could not be explained [see Fig. 5(e)].

## V. CONCLUSIONS

To conclude, the magneto-optical response of a series of p-type HgTe/CdHgTe quantum wells, with normal as well as inverted band ordering, has been comprehensively studied in the THz and infrared spectral ranges. We have found that the observed response cannot be explained within the standard 4-band model, which is nowadays widely applied and which assumes the full axial symmetry of the system (along the growth axis). We propose that additionally observed spectral features (avoided crossing of transitions and "forbidden" lines) have their origin in specific symmetry lowering effects. These mainly include the bulk inversion asymmetry and anisotropy of chemical bonds at the heterointerfaces. Our calculations of the magneto-optical response with both these asymmetries properly included in the corresponding Hamiltonian support this conclusion.

## ACKNOWLEDGMENTS

This work was supported by the Russian Science Foundation (Grant No. 16-12-10317), by CNRS through the LIA TeraMIR project and by ANR through Dirac3D project. The authors thank E. L. Ivchenko for fruitful discussions IIA effects. L.B. also acknowledges Vernadski scholarship of the French government.

*Supplemental Material:* **Landau level spectroscopy of valence bands in HgTe quantum wells: Effects of symmetry lowering**

L. S. Bovkun,[1,2,*] A. V. Ikonnikov,[1,3] V. Ya. Aleshkin,[1,4] K. E. Spirin,[1] V. I. Gavrilenko,[1,4] N. N. Mikhailov,[5] S. A. Dvoretskii,[5] F. Teppe,[6] B.A. Piot,[2] M. Potemski,[2] M. Orlita[2,7,†]

[1]*Institute for Physics of Microstructures RAS, 603950 Nizhny Novgorod, Russia*
[2]*Laboratoire National des Champs Magnétiques Intenses, LNCMI-CNRS-UGA-UPS-INSA-EMFL, 38042 Grenoble, France*
[3]*Faculty of Physics, M.V. Lomonosov Moscow State University, 119991 Moscow, Russia*
[4]*Lobachevsky State University of Nizhny Novgorod, 603950 Nizhny Novgorod, Russia*
[5]*Institute of Semiconductor Physics, Siberian Branch RAS, 630090 Novosibirsk, Russia*
[6]*Laboratoire Charles Coulomb (L2C), UMR CNRS 5221, GIS-TERALAB, Université Montpellier II, F-34095 Montpellier, France*
[7]*Institute of Physics, Charles University, 12116 Prague, Czech Republic*
(Dated)

### Effects of the Structure Inversion Asymmetry

In the present work, we do not consider possible effects of the Structure Inversion Asymmetry, i.e. we consider our QWs to be symmetric, implying strictly rectangular potential. The latter is partially justified by a good agreement between experimental and theoretical results for optical transitions between adjacent but also further lying subbands. A giant (up to 30 meV) conduction band splitting (Rashba splitting) resulting, in particular, in CR line splitting (in classic magnetic fields) up to 10% [1,2] are known to appear in single-side selectively doped HgTe/CdHgTe QWs with an inverted band structure at electron concentrations over $10^{12}$ cm$^{-2}$. Such conditions are very far from those in our samples, which are basically symmetric and without any intentional doping (with hole density close to $10^{11}$ cm$^{-2}$).

Let us note, however, that the EDSR lines invoked by Referee A are in principle present even without Rashba coupling. As a matter of fact, the fairly complex structure of the Hamiltonian, which lacks both inversion and axial symmetry gives rise to basically all possible electric-dipole excitations, including those changing the spin projection of electrons (EDSR). In our previous work [3], we find out that «the oscillator strength of the EDSR transition line is five orders of magnitude lower than those for α− and β transitions» in the rectangular HgTe/CdHgTe (013) QW with the normal band structure for the electric field along x // [100].

Table A. Squares of matrix elements of optical transitions between Landau levels in the axial approximation. Results are given at a magnetic induction of 6T for QW with *d*=5nm.

|  | rectangular | | trapezoidal | |
| --- | --- | --- | --- | --- |
|  | $|x_{if}|^2$ | $|y_{if}|^2$ | $|x_{if}|^2$ | $|y_{if}|^2$ |
| α ($0_\uparrow \to 1_\uparrow$) | 3671 | 3862 | 3828 | 4050 |
| β ($-2_\downarrow \to -1_\downarrow$) | 1416 | 1317 | 1077 | 1015 |
| ESDR ($0_\uparrow \to -1_\downarrow$) | $10^{-8}$ | 54 | $10^{-2}$ | 41 |
| α / ESDR ratio | >$10^5$ | ~72 | >$10^5$ | ~99 |

Recently we repeated the calculations (see Table A) for unpolarized radiation and get ESDR matrix element of 54 solely to $|y_{if}|^2$ (that were neglected in [3]) in again rectangular QW 5 nm wide. The next step was to calculate EDSR for QW of trapezium shape with one inclined heterointerface 1 nm wide. In this case for the electric field along x//[100], $|x_{if}|^2$ for EDSR transition $0\uparrow \to -1\downarrow$ thought increased up to $10^{-2}$ proved to be still negligible while the other above figures were practically the same. Therefore, for unpolarized radiation, the probability of EDSR transition $0\uparrow \to -1\downarrow$ is at least 10 times smaller than other dominant absorption lines. So, EDSR intensity seems to be insensitive to HgTe/CdHgTe QW shape and small compared to CR one that explains its absence in the magnetoabsorption spectra.

### An anisotropic Hamiltonian term describing the lack of the inversion symmetry in the bulk crystal (BIA)

The anisotropic part of the total Hamiltonian $H_a$ includes terms describing the lack of the inversion symmetry in the bulk crystal (BIA) and the symmetry lowering at the heterointerfaces (IIA). The BIA term was derived from 14×14 Kane Hamiltonian proposed in Ref. 4, from which also the corresponding parameters were taken (same for CdTe and HgTe). In our approach we imply set of wave function from Ref. 5 using following conversion formulae:

$$H_{i',j'} = H^t_{i',j'} - \sum_{n=1} \frac{H^t_{i',n} H^t_{n,j'}}{E_n} = H^t_{i',j'} + H^{BIA},$$

where $H^t$ stands for 14×14 Kane Hamiltonian and indexes *i'* and *j'* go through states of $\Gamma_{7v}$, $\Gamma_{8v}$, $\Gamma_{6c}$, while index n travels within states of $\Gamma_{8c}$ and $\Gamma_{7c}$, $E_n$ – energies of $\Gamma_{8c}$ and $\Gamma_{7c \text{ band edges}}$ at the center of Brillouin zone, $H^{BIA}$ implicates bulk inversion asymmetry effects. Components of $H^{BIA}$ are given by the following expressions.

$$H^{BIA}_{1,1} = H^{BIA}_{1,2} = H^{BIA}_{2,2} = 0$$

$$H_{1,3}^{BIA} = \sqrt{2}\left[(0.8A+0.4B)k_-k_z + i(0.15A+0.05B)k_-k_+ - i(0.225A+0.15B)k_-^2 - i(0.3A+0.1B)k_z^2 + i0.075Ak_+^2\right]$$

$$H_{1,4}^{BIA} = \sqrt{\frac{2}{3}}\left[0.4Ak_+^2 - 0.4(A+B)k_-^2 + i(0.6A+0.15B)k_+k_z + i(0.6A+0.45B)k_-k_z\right]$$

$$H_{1,5}^{BIA} = \sqrt{\frac{2}{3}}\left[(0.8A+0.4B)k_+k_z - i0.15(A+B)k_-k_+ - i(0.075A+0.15B)k_-^2 + i0.225Ak_+^2 + i0.3(A+B)k_z^2\right]$$

$$H_{1,6}^{BIA} = \frac{\sqrt{2}}{3}\left[0.8Bk_z^2 - 0.4Bk_-k_+ - i0.45Bk_-k_z + i0.45Bk_+k_z\right]$$

$$H_{1,7}^{BIA} = \frac{1}{\sqrt{3}}\left[0.4Ak_-^2 - i0.6Ak_-k_z - 0.4Ak_+^2 - i0.6Ak_+k_z + 3Zk_z\right]$$

$$H_{1,8}^{BIA} = \frac{1}{\sqrt{3}}\left[3Zk_- - i0.45Ak_+^2 - 1.6Ak_+k_z - i0.6Ak_z^2 + i0.15Ak_-^2 + i0.3Ak_-k_+\right]$$

$$H_{2,3}^{BIA} = -H_{1,6}^{BIA}$$

$$H_{2,4}^{BIA} = (H_{1,5}^{BIA})^*$$

$$H_{2,5}^{BIA} = -(H_{1,4}^{BIA})^*$$

$$H_{2,6}^{BIA} = (H_{1,3}^{BIA})^*$$

$$H_{2,7}^{BIA} = \frac{1}{\sqrt{3}}\left[3Zk_+ + 1.6Ak_-k_z - i0.6Ak_z^2 - i0.45Ak_-^2 + i0.3Ak_-k_+ + i0.15Ak_+^2\right]$$

$$H_{2,8}^{BIA} = -H_{1,7}^{BIA}$$

$$H_{3,3}^{BIA} = H_{4,4}^{BIA} = H_{5,5}^{BIA} = H_{6,6}^{BIA} = D$$

$$H_{3,4}^{BIA} = H_{3,5}^{BIA} = H_{3,6}^{BIA} = H_{4,5}^{BIA} = H_{4,6}^{BIA} = H_{5,6}^{BIA} = H_{7,8}^{BIA} = 0$$

$$H_{3,7}^{BIA} = \sqrt{6}\left[0.8Sk_+ + i0.6Sk_z - i0.12G(k_+^2+k_-^2) + 0.64Gk_-k_z + 0.36Gk_+k_z - i0.48Gk_z^2 + i0.24Gk_-k_+\right]$$

$$H_{3,8}^{BIA} = \sqrt{6}\left[1.6Sk_z - i0.3Sk_- + i0.9Sk_+ + 0.09Gk_-^2 - 0.18Gk_-k_+ - 0.91Gk_+^2 - i0.48Gk_+k_z + 0.36Gk_z^2 + i0.48Gk_-k_z\right]$$

$$H_{4,7}^{BIA} = \sqrt{2}\left[-i0.9S(k_-+k_+) + 0.92Gk_z^2 - 0.46Gk_-k_+ + i1.44Gk_z(k_+-k_-) - 0.27G(k_+^2+k_-^2)\right]$$

$$H_{4,8}^{BIA} = -\sqrt{3}H_{3,7}^{BIA}$$

$$H_{5,7}^{BIA} = \sqrt{2}\left[2.4Sk_- - i1.8Sk_z - i0.36G(k_+^2+k_-^2) - 1.08Gk_-k_z - 1.92Gk_+k_z - i1.44Gk_z^2 + i0.72Gk_-k_+\right]$$

$$H_{5,8}^{BIA} = -H_{4,7}^{BIA}$$

$$H_{6,7}^{BIA} = \sqrt{6}\left[i0.3Sk_+ - i0.9Sk_- + 0.91Gk_-^2 + 0.18Gk_-k_+ - 0.09Gk_+^2 - 0.36Gk_z^2 + i0.48Gk_z(k_+-k_-)\right]$$

$$H_{6,8}^{BIA} = \sqrt{6}\left[-0.8Sk_- + i0.6Sk_z - i0.24Gk_-k_+ + 0.36Gk_-k_z + 0.64Gk_+k_z + i0.12G(k_+^2+k_-^2) + i0.48Gk_z^2\right]$$

$$H_{7,7}^{BIA} = H_{8,8}^{BIA} = D_1,$$

where $k_\pm = k_x \pm ik_y$ and

$$A = -i\frac{P'Q}{E_0+\Delta_0}, \quad B = -i\frac{P'Q\Delta_0}{E_0(E_0+\Delta_0)}, \quad Z = \frac{P'\Delta^-(3E_0+2\Delta_0)}{9E_0(E_0+\Delta_0)}, \quad S = -i\frac{Q\Delta^-(3E_0+2\Delta_0)}{18E_0(E_0+\Delta_0)},$$

$$G = \frac{Q^2\Delta_0}{6E_0(E_0+\Delta_0)}, \quad D = -\frac{|\Delta^-|^2}{9(E_0+\Delta_0)}, \quad D_1 = -\frac{4}{9}\frac{|\Delta^-|^2}{E_0} + \frac{Q^2\Delta_0}{3E_0(E_0+\Delta_0)}k^2,$$

Where letter designation are taken from [4]. The rest of the components are derived as Hermitian conjugation of corresponding elements. The magnetic field is taken into account via Zeeman term and substitution (cf. Ref. 6) $k_+ = \frac{\sqrt{2}}{\lambda} a^+$, $k_- = \frac{\sqrt{2}}{\lambda} a$, $\lambda = \sqrt{\frac{\hbar c}{|eB|}}$, where $a^+$ and $a$ are creation and annihilation operators.

### Matrix elements for optical transitions between Landau levels

Matrix elements for optical transitions between Landau levels were calculated for unpolarized radiation. The probability of such a transition is proportional to the square of the dipole moment matrix element (product of the electron charge on the matrix element of the coordinate operator). The matrix element $r_{f,i}$ of the coordinate operator between the initial $|i\rangle$ and final $|f\rangle$ states satisfies the equation:

$$v_{f,i} = \frac{i}{\hbar}[H, r]_{f,i} = \frac{i}{\hbar}(\varepsilon_f - \varepsilon_i) r_{f,i},$$

where $v$ is the velocity operator. First, one has to find the velocity operator, then calculate its matrix element and finally find the matrix element of the coordinate operator from the equation $H = H_s + H_a$. Components of velocity operator in magnetic field are given by:

$$v_x = \begin{pmatrix}
T_x & 0 & -\frac{1}{\sqrt{2\hbar}}P & 0 & \frac{1}{\sqrt{6\hbar}}P & 0 & 0 & -\frac{1}{\sqrt{3\hbar}}P \\
0 & T_x & 0 & -\frac{1}{\sqrt{6\hbar}}P & 0 & \frac{1}{\sqrt{2\hbar}}P & -\frac{1}{\sqrt{3\hbar}}P & 0 \\
-\frac{1}{\sqrt{2\hbar}}P & 0 & U_x + V_x & S_x + s_x & R_x & 0 & -\frac{(S_x + s_x)}{\sqrt{2}} & -\sqrt{2}R_x \\
0 & -\frac{1}{\sqrt{6\hbar}}P & S_x^+ + s_x^+ & U_x - V_x & C_x & R_x & \sqrt{2}V_x & \sqrt{\frac{3}{2}}Z_x \\
\frac{1}{\sqrt{6\hbar}}P & 0 & R_x^+ & C_x^+ & U_x - V_x & -S_x + s_x & \sqrt{\frac{3}{2}}Z_x & -\sqrt{2}V_x \\
0 & \frac{1}{\sqrt{2\hbar}}P & 0 & R_x^+ & -S_x^+ + s_x^+ & U_x + V_x & \sqrt{2}R_x^+ & -\frac{(S_x^+ - s_x^+)}{\sqrt{2}} \\
0 & -\frac{1}{\sqrt{3\hbar}}P & -\frac{(S_x^+ + s_x^+)}{\sqrt{2}} & \sqrt{2}V_x & \sqrt{\frac{3}{2}}Z_x^+ & \sqrt{2}R_x & U_x & C_x \\
-\frac{1}{\sqrt{3\hbar}}P & 0 & -\sqrt{2}R_x^+ & \sqrt{\frac{3}{2}}Z_x^+ & -\sqrt{2}V_x & -\frac{(S_x - s_x)}{\sqrt{2}} & C_x^+ & U_x
\end{pmatrix}$$

$$v_y = \begin{pmatrix}
T_y & 0 & -\frac{i}{\sqrt{2\hbar}}P & 0 & \frac{-i}{\sqrt{6\hbar}}P & 0 & 0 & \frac{i}{\sqrt{3\hbar}}P \\
0 & T_y & 0 & -\frac{i}{\sqrt{6\hbar}}P & 0 & \frac{-i}{\sqrt{2\hbar}}P & -\frac{i}{\sqrt{3\hbar}}P & 0 \\
\frac{i}{\sqrt{2\hbar}}P & 0 & U_y + V_y & S_{y0} + s_y & R_{y-} & 0 & -\frac{(S_y + s_y)}{\sqrt{2}} & -\sqrt{2}R_y \\
0 & \frac{i}{\sqrt{6\hbar}}P & S_y^+ + s_y^+ & U_y - V_y & C_y & R_y & \sqrt{2}V_y & \sqrt{\frac{3}{2}}Z_y \\
\frac{i}{\sqrt{6\hbar}}P & 0 & R_y^+ & C_y^+ & U_y - V_y & -S_y + s_y & \sqrt{\frac{3}{2}}Z_y & -\sqrt{2}V_y \\
0 & \frac{i}{\sqrt{2\hbar}}P & 0 & R_y^+ & -S_y^+ + s_y^+ & U_y + V_y & \sqrt{2}R_y^+ & -\frac{(S_y^+ - s_y^+)}{\sqrt{2}} \\
0 & \frac{i}{\sqrt{3\hbar}}P & -\frac{(S_y^+ + s_y^+)}{\sqrt{2}} & \sqrt{2}V_y & \sqrt{\frac{3}{2}}Z_y^+ & \sqrt{2}R_y & U_y & C_y \\
-\frac{i}{\sqrt{3\hbar}}P & 0 & -\sqrt{2}R_y^+ & \sqrt{\frac{3}{2}}Z_y^+ & -\sqrt{2}V_y & -\frac{(S_y - s_y)}{\sqrt{2}} & C_y^+ & U_y
\end{pmatrix}$$

where

$$T_x = \frac{\hbar}{m_0 \lambda \sqrt{2}} (2F + 1)(a^+ + a), \qquad T_y = \frac{i\hbar}{m_0 \lambda \sqrt{2}} (2F + 1)(a - a^+)$$

$$U_x = -\frac{\hbar}{m_0\lambda\sqrt{2}}\gamma_1(a+a^+), \quad U_y = -\frac{i\hbar}{m_0\lambda\sqrt{2}}\gamma_1(a-a^+)$$

$$V_x = -\frac{\hbar}{m_0\lambda\sqrt{2}}\gamma_2(a+a^+),$$

$$V_y = -i\frac{\hbar}{m_0\lambda\sqrt{2}}(\gamma_2 - 0.54(\gamma_2-\gamma_3))(a-a^+) - \frac{\hbar}{m_0}0.36\{(\gamma_2-\gamma_3),k_z\}$$

$$S_x = \frac{\hbar\sqrt{3}}{2m_0}\{\gamma_3,k_z\},$$

$$S_y = -i\frac{\hbar\sqrt{3}}{2\sqrt{2}m_0\lambda}0.96i(\gamma_2-\gamma_3)(a-a^+) - i\frac{\hbar\sqrt{3}}{2m_0}\left(0.36\{(\gamma_2-\gamma_3),k_z\}+\{\gamma_3,k_z\}\right)$$

$$s_x = \frac{\hbar}{2m_0}\sqrt{3}[\kappa,k_z], \quad s_y = \frac{-i\hbar}{2m_0}\sqrt{3}[\kappa,k_z]$$

$$R_x = \frac{\hbar}{m_0\lambda}\sqrt{\frac{3}{2}}\left(2\gamma_2 a + (\gamma_2-\gamma_3)(a^+ - a)\right),$$

$$R_y = \frac{i\hbar}{m_0\lambda}\sqrt{\frac{3}{2}}\left(-2\gamma_2 a + (\gamma_2-\gamma_3)(a+a^+) + 0.18(\gamma_2-\gamma_3)(a-a^+)\right) - 0.12\frac{\hbar}{m_0}\sqrt{3}\{(\gamma_2-\gamma_3),k_z\}$$

$$C_x = \frac{\hbar}{m_0}[\kappa,k_z], \quad C_y = \frac{-i\hbar}{m_0}[\kappa,k_z],$$

$$Z_x = \frac{\hbar\sqrt{3}}{2m_0}\left(\{\gamma_3 k_z\} - \frac{1}{3}[\kappa,k_z]\right),$$

$$Z_y = -i\frac{\hbar\sqrt{3}}{2m_0}\left(0.36\{(\gamma_2-\gamma_3),k_z\} + \{\gamma_3 k_z\} + \frac{1}{3}[\kappa,k_z]\right) + 0.96\frac{\hbar}{2m_0\lambda}\sqrt{\frac{3}{2}}(\gamma_2-\gamma_3)(a-a^+)$$

$$z_x = \frac{\hbar\sqrt{3}}{2m_0}\left(\{\gamma_3,k_z\} - \frac{1}{3}[\kappa,k_z]\right),$$

$$z_y = i\frac{\hbar\sqrt{3}}{2m_0}\left(\{\gamma_3,k_z\} - \frac{1}{3}[\kappa,k_z] + 0.36\{(\gamma_2-\gamma_3),k_z\}\right) - 0.96\frac{\hbar}{2m_0\lambda}\sqrt{\frac{3}{2}}(\gamma_2-\gamma_3)(a-a^+)$$

---

\* bovkun@ipmras.ru  
† milan.orlita@lncmi.cnrs.fr